\documentclass[prd,twocolumn,groupedaddress,showpacs,nofootinbib]{revtex4}
\usepackage{graphicx}
\usepackage{dcolumn}
\usepackage{bm}
\usepackage{amssymb}
\usepackage{mathrsfs}
\usepackage{amsmath}
\usepackage{epsfig}
\usepackage[dvips]{color}
\usepackage{hhline}

\begin{document}

\title{A Left-Right Symmetric Model for Neutrino Masses, Baryon Asymmetry and Dark Matter}

\author{Pei-Hong Gu}
\email{peihong.gu@mpi-hd.mpg.de}

\affiliation{Max-Plank-Institut f\"{u}r Kernphysik, Saupfercheckweg
1, 69117 Heidelberg, Germany}

\begin{abstract}

In the left-right symmetric models without bi-doublet Higgs scalars,
the standard model fermions can obtain masses by integrating out
heavy charged singlet fermions. We find the decays of heavy neutral
singlet fermions, responsible for generating small neutrino masses,
can simultaneously produce a left-handed lepton asymmetry for baryon
asymmetry and a relic density of right-handed neutrinos for dark
matter. Benefited from the left-right symmetry, the properties of
the dark matter can be related to the generation of the neutrino
masses and the baryon asymmetry. We also indicate that the decays of
the non-thermally produced right-handed neutrinos can explain the
observed fluxes of 511 keV photons from the Galactic bulge.

\end{abstract}


\maketitle

\section{Introduction}

The $SU(3)_c^{}\times SU(2)_{L}^{}\times U(1)_{Y}^{}$ standard model
(SM) has been tested to a very high accuracy, but it has been
suffering big challenges from particle physics and cosmology. For
example, it can't explain the existence of non-baryonic dark matter,
which has been confirmed by precisely cosmological observations
\cite{dunkley2008},
\begin{eqnarray}
\label{dmatter} \Omega_\chi^{}h^2_{}=0.1099\pm0.0062\,.
\end{eqnarray}
Here $h=0.719^{+0.026}_{-0.027}$ \cite{dunkley2008}. So far we know
little on the true identity of the dark matter although there have
been many interesting candidates. The cosmological observations also
indicates that the present universe doesn't contain significant
amount of baryonic antimatter. This baryon asymmetry again requires
supplementing new ingredients to the existing theory. The density of
the baryonic matter is measured by \cite{dunkley2008}
\begin{eqnarray}
\label{bmatter} \Omega_{b}^{}h^2_{}=0.02273\pm0.00062\,,
\end{eqnarray}
which is intriguingly comparable to that of the dark matter. The
coincidence between the dark and baryonic matter implies that they
may have a specifical relation \cite{kuzmin1997} although their
creation and evolution are usually understood by unrelated
mechanisms. Furthermore, observations on solar, atmospheric, reactor
and accelerator neutrino oscillations have established the
phenomenon of massive and mixing neutrinos \cite{stv2008},
\begin{subequations}
\begin{eqnarray}
\Delta m_{21}^2 &=&7.65^{+0.23}_{-0.20} \times
10^{-5}_{}\,\textrm{eV}^2_{}\,,\\
\vspace{10mm} \Delta m_{31}^2 &=& \pm 2.4^{+0.12}_{-0.11}\times
10^{-3}_{}\,\textrm{eV}^2_{}\,,\\
\sin^2_{}\theta_{23}^{} &=& 0.50^{+0.07}_{-0.06} \,,\\
\sin^2_{}\theta_{12}^{} &=& 0.304^{+0.022}_{-0.016}\,, \\
\sin^2_{}\theta_{13}^{} &=&0.01^{+0.016}_{-0.011}\,,
\end{eqnarray}
\end{subequations}
which is well consistent with the cosmological limits on the sum of
the neutrino masses \cite{dunkley2008},
\begin{eqnarray}
\Sigma m_i^{}<1.3\,\textrm{eV}\quad (95\%\,\textrm{CL})\,.
\end{eqnarray}
The smallness of the neutrino masses inspires a seesaw
\cite{minkowski1977,yanagida1979,grs1979,glashow1980,ms1980}
extension of the SM. The seesaw models
\cite{minkowski1977,yanagida1979,grs1979,glashow1980,ms1980,mw1980}
also allow the attractive leptogenesis
\cite{fy1986,lpy1986,fps1995,pilaftsis1997,ms1998,bpy2005} mechanism
for generating the observed baryon asymmetry.

On the other hand, new physics beyond the SM is also motivated by
some theoretical considerations. For example, the SM accommodates
the left- and right-handed fermions in different ways: the
left-handed fermions transform as doublets but the right-handed ones
transform as singlets. The unequal treatment on the left- and
right-handed fermions means parity violation. If we start with a
theory with left-right symmetry \cite{ps1974} at high energy, where
the left- and right-handed fermions are both placed in doublets and
the parity is conserved, the spontaneously left-right symmetry
breaking can induce the parity violation as observed at low energy.
The left-right symmetric models are based on the gauge group
$SU(3)_c^{}\times SU(2)_{L}^{}\times SU(2)_{R}^{}\times
U(1)_{B-L}^{}$. Within this context, the electric charge ($Q$) can
be well defined in terms of lepton ($L$) and baryon ($B$) numbers,
\begin{eqnarray}
Q= T_{L_3}^{}+T_{R_3}^{}+\frac{B-L}{2}\,,
\end{eqnarray}
where $\overrightarrow{T}_{L}^{}$, $\overrightarrow{T}_{R}^{}$ and
$B-L$ are the generators of $SU(2)_{L}^{}$, $SU(2)_{R}^{}$ and
$U(1)_{B-L}^{}$, respectively. This provides us a natural
explanation for the choice of hypercharge charge in the SM. An
important prediction of the left-right symmetry is the existence of
right-handed neutrinos, which are trivial and hence absent in the
SM. In the most popular left-right symmetric models with bi-doublet
and triplet Higgs scalars \cite{ms1980}, the right-handed neutrinos
can naturally obtain heavy Majorana masses to realize the seesaw and
the leptogenesis after the left-right symmetry breaking.

In this paper we consider the left-right symmetric models without
bi-doublet Higgs scalars \cite{bmu2003}. If the Higgs content only
contains a left-handed doublet and its right-handed partner, one
\cite{siringo2003} need assume the Higgs potential to be $O(2)$
invariant for giving an acceptable symmetry breaking pattern.
Alternatively we shall extend the original model \cite{bmu2003} with
spontaneous D-parity violation \cite{cmp1984}. For generating the
masses of the SM fermions, we introduce some heavy charged singlet
fermions with Dirac masses and then integrating them out. For the
neutrinos, we introduce left- and right-handed neutral singlet
fermions with not only Dirac but also Majorana masses. The
right(left)-handed neutral singlets can dominate the seesaw for the
left(right)-handed neutrino masses. Their decays can produce the
observed baryon asymmetry through the leptogenesis and can give the
right-handed neutrinos a desired relic density for the dark matter.
In this left-right symmetric scenario, the properties of the dark
matter are related to the parameters for generating the neutrino
masses and the baryon asymmetry.

\section{The model}

The most general scalar potential of our model is
\begin{eqnarray}
\label{potential}
V&=&-\mu_1^2\sigma^2_{}-\mu^2_{2}\left(\phi_L^\dagger\phi_L^{}+\phi_R^\dagger\phi_R^{}\right)
+\mu_3^{}\sigma\left(\phi_L^\dagger\phi_L^{}\right.\nonumber\\
&&\left.-\phi_R^\dagger\phi_R^{}\right)+\lambda_1^{}\sigma^4_{}
+\lambda_2^{}\left[\left(\phi_L^\dagger\phi_L^{}\right)^2_{}+\left(\phi_R^\dagger\phi_R^{}\right)^2_{}\right]
\nonumber\\
&&+\lambda_3^{}\phi^\dagger_L\phi^{}_L
\phi^\dagger_R\phi_R^{}+\lambda_4^{}\sigma^2_{}\left(\phi^\dagger_L\phi_L^{}+\phi^\dagger_R\phi^{}_R\right)\,,
\end{eqnarray}
with $\sigma \rightarrow -\sigma$ and $\phi_L^{} \rightarrow
\phi_R^{}$ under the parity symmetry. Here $\lambda_{1,2}^{}>0$,
$\lambda_3^{}>-2\lambda_2^{}$ and
$\lambda_4^{}>-2\sqrt{\lambda_1^{}\lambda_2^{}}$ so that the
potential is bounded from below. For appropriate parameter choice,
it is easy to give the desired vacuum expectation values (VEVs),
\begin{eqnarray}
\label{vev} v_R^{}=\langle\phi_R^{}\rangle&\gg&
v_L^{}=\langle\phi_L^{}\rangle\simeq 174\,\textrm{GeV}\,.
\end{eqnarray}
In the present model without the Higgs bi-doublets, the charged
gauge bosons $W_L^\pm$ and $W_R^\pm$ are mass eigenstates
\cite{siringo2003},
\begin{eqnarray}
\label{wmass} m_{W_L^{}}^{}=\frac{1}{\sqrt{2}}g v_L^{}\,,\quad
m_{W_R^{}}^{}=\frac{1}{\sqrt{2}}g v_R^{}\,.
\end{eqnarray}
The left- and right-handed fermions in the SM with the right-handed
neutrinos are placed in doublets for each family. For generating the
masses of the SM fermions, we introduce some heavy charged singlets
to construct the Yukawa couplings to the Higgs and fermion doublets
so that we can derive the SM yukawa couplings by integrating out
these singlets \cite{bmu2003},
\begin{eqnarray}
\label{lagrangian1a} \mathcal{L}&\supset&
-y_D^{}\left(\bar{q}_L^{}\phi_L^{}D_R^{}+\bar{q}_R^{}\phi_R^{}D_L^{}\right)-M_D^{}\bar{D}_L^{}D_R^{}\nonumber\\
&&-y_U^{}\left(\bar{q}_L^{}\tilde{\phi}_L^{}U_R^{}+\bar{q}_R^{}\tilde{\phi}_R^{}U_L^{}\right)-M_U^{}\bar{U}_L^{}U_R^{}\nonumber\\
&&-y_E^{}\left(\bar{l}_L^{}\phi_L^{}E_R^{}+\bar{l}_R^{}\phi_R^{}E_L^{}\right)-M_E^{}\bar{E}_L^{}E_R^{}\nonumber\\
&&+\textrm{H.c.}\nonumber \\
&\Rightarrow&
-y_d^{}\bar{q}_L^{}\phi_L^{}d_R^{}-y_u^{}\bar{q}_L^{}\tilde{\phi}_L^{}u_R^{}-y_e^{}\bar{l}_L^{}\phi_L^{}e_R^{}\nonumber\\
&&+\textrm{H.c.}\,,
\end{eqnarray}
where the SM Yukawa couplings are given by
\begin{subequations}
\begin{eqnarray}
\label{smyd}
y_d^{}&=&-y_D^{L}\frac{v_R^{}}{M_D^{}}y_D^{R\dagger}\,,\\
\label{smyu}
y_u^{}&=&-y_U^{L}\frac{v_R^{}}{M_U^{}}y_U^{R\dagger}\,,\\
\label{smye} y_e^{}&=&-y_E^{L}\frac{v_R^{}}{M_E^{}}y_E^{R\dagger}\,.
\end{eqnarray}
\end{subequations}
Here we have chosen the base where the mass matrices $M_{D,U,E}^{}$
are real and diagonal. Note the mass matrices $M_{D,U,E}^{}$ now are
not hermitian as a result of the spontaneous D-parity violation,
i.e.
\begin{eqnarray}
M_{D,U,E}^{}&=&M_{D,U,E}^{0}+h_{D,U,E}^{}\langle\sigma\rangle
\end{eqnarray}
with $M_{D,U,E}^0=M_{D,U,E}^{0\dagger}$ while
$h_{D,U,E}^{\dagger}=-h_{D,U,E}^{}$. If the Yukawa couplings
$h_{D,U,E}^{}$ vanish, the left-handed yukawa couplings
$y_{D,U,E}^{L}$ should be equal to the right-handed ones
$y_{D,U,E}^{R}$ as $M_{D,U,E}^{0}$ are hermitian matrices. Since the
Yukawa coupling of top quark is very close to $1$, Eq. (\ref{smyu})
will constrain $M_T^{}\lesssim v_R^{}$ for $y_T^{}< \sqrt{4\pi}$.
For simplicity, we shall conveniently assume $M_D^{}=M_U^{}=M_E^{}=
v_R^{}$ to make the heavy charged singlets decoupled from the
following discussions.

In the neutrino sector, we consider the left- and right-handed
neutral singlets with the Yukawa couplings and the masses as below,
\begin{eqnarray}
\label{lagrangian1b}
\mathcal{L}&\supset&-y_N^{}\left(\bar{l}_L^{}\tilde{\phi}_L^{}N_R^{}+\bar{l}_R^{}\tilde{\phi}_R^{}N_L^{}\right)
-M_N^{D}\bar{N}_L^{}N_R^{}\nonumber\\
&&-\frac{1}{2}M_N^M \left( \bar{N}_L^c N_L^{}+ \bar{N}_R^c
N_R^{}\right)+\textrm{H.c.}\,.
\end{eqnarray}
In the presence of the spontaneous D-parity violation, the Dirac
mass matric $M_N^D$ is not hermitian, i.e.
\begin{eqnarray}
M_{N}^{D}&=&M_{N}^{0}+h_{N}^{}\langle\sigma\rangle
\end{eqnarray}
with $M_{N}^0=M_{N}^{0\dagger}$ while $h_N^{\dagger}=-h_N^{}$.
Furthermore, we have forbidden the Yukawa couplings
$\bar{l}_L^{}\tilde{\phi}_L^{}N_L^{c}$,
$\bar{l}_R^{}\tilde{\phi}_R^{}N_R^c$ and their CP conjugates. This
can be achieved by imposing a discrete, global or local symmetry.
For example, we consider a $U(1)_X^{}$ local symmetry under which
$D_{L,R}^{},U_{L,R}^c,E_{L,R}^{},N_{L,R}^c,\phi_{L,R}^{\ast}$ carry
a quantum number $X=1$. Clearly, this $U(1)_X^{}$ is free of gauge
anomaly. In this context, the Yukawa couplings and the Dirac mass
terms in Eqs. (\ref{lagrangian1a}) and (\ref{lagrangian1b}) are
allowed while the Majorana mass terms in Eq. (\ref{lagrangian1b})
are forbidden. To break this $U(1)_X^{}$ \footnote{There is a
kinetic mixing between the gauge bosons associated with $U(1)_X^{}$
and $U(1)_{B-L}^{}$. We simply assume this mixing is small enough to
fulfill the experimental constraints.}, we can introduce a singlet
scalar $\xi$ with Yukawa couplings to the neutral singlets
$N_{L,R}^{}$,
\begin{eqnarray}
\mathcal{L}&\supset& -\frac{1}{2}f_N^{} \left( \xi\bar{N}_L^c
N_L^{}+\xi^\ast_{} \bar{N}_R^c N_R^{}\right)+\textrm{H.c.}\,.
\end{eqnarray}
Through the above Yukawa interactions, the Majorana masses in Eq.
(\ref{lagrangian1b}) can be given by
\begin{eqnarray}
M_N^M=f_N^{} \langle\xi \rangle\,.
\end{eqnarray}
By integrating out the neutral singlets, the full neutrino masses
would contain a Dirac mass term and two Majorana ones,
\begin{eqnarray}
\label{lagrangian2} \mathcal{L}\supset-\frac{1}{2}\bar{\nu}_L^{}
m_L^{} \nu_L^c-\frac{1}{2}\bar{\nu}_R^{} m_R^{}
\nu_R^c-\bar{\nu}_L^{} m_D^{} \nu_R^{}+\textrm{H.c.}
\end{eqnarray}
with
\begin{subequations}
\begin{eqnarray}
\label{massl}
m_L^{}&=&-y_N^{}\frac{1}{M_N^M}y_N^T v_L^2\,,\\
\label{massr}
m_R^{}&=&-y_N^{}\frac{1}{M_N^M}y_N^T v_R^2\,,\\
\label{massd} m_D^{}&=&y_N^{}\frac{1}{M_N^M}(M_N^{D})^T_{}
\frac{1}{M_N^M}y_N^\dagger v_L^{}v_R^{}\,.
\end{eqnarray}
\end{subequations}
Here we have assumed
\begin{eqnarray}
\label{assumption1} M_N^M\gg M_N^D,y_N^{}v_R^{},y_N^{}v_L^{}\,,
\end{eqnarray}
and then have defined the left- and right-handed Majorana fermions,
\begin{subequations}
\begin{eqnarray}
L_i^{}&=&N_{L_i^{}}^{}+N_{L_i^{}}^c\,,\\
R_i^{}&=&N_{R_i^{}}^{}+N_{R_i^{}}^c\,,
\end{eqnarray}
\end{subequations}
by choosing the base where the Majorana mass matrix $M_N^M$ is real
and diagonal,
\begin{eqnarray}
M_N^M=\textrm{diag}\{M_1^{},M_2^{},M_3^{}\}\simeq M.
\end{eqnarray}
Clearly, the right-handed neutrinos will give their left-handed
partners an additional Majorana mass term through the seesaw since
their Dirac masses are not vanishing. This gift is negligible,
\begin{eqnarray}
\delta
m_L^{}=-m_D^{}\frac{1}{m_R^{\dagger}}m_D^T=\mathcal{O}\left[\left(\frac{M_N^D}{M_N^M}\right)^2_{}\right]m_L^{}\ll
m_L^{}\,.
\end{eqnarray}
Therefore, we can well define the left- and right-handed Majorana
neutrinos,
\begin{subequations}
\begin{eqnarray}
\nu &=&\nu_L^{}+\nu_L^c\,,\\
\chi &=&\nu_R^{}+\nu_R^c\,,
\end{eqnarray}
\end{subequations}
which have mass matrices with a same texture,
\begin{subequations}
\label{masslr}
\begin{eqnarray}
m_\nu^{}&=&m_L^{}=U^\dagger_{}\textrm{diag}\{m_1^{},m_2^{},m_3^{}\}U^\ast_{}\,,\\
m_\chi^{}&=&m_R^{}=U^\dagger_{}\textrm{diag}\{m_1^{},m_2^{},m_3^{}\}U^\ast_{}\frac{v_R^2}{v_L^2}\,.
\end{eqnarray}
\end{subequations}
Here $U$ is the MNS lepton flavor mixing matrix \cite{mns1962}.

\section{Baryon asymmetry}

We now discuss the physics below the scale $M_N^M$, which is assumed
much smaller than the left-right symmetry breaking scale $v_R^{}$,
i.e.
\begin{eqnarray}
\label{assumption2} M_N^M \ll v_R^{}\,.
\end{eqnarray}
We thus integrate out the right-handed physical Higgs boson and then
derive the following Lagrangian,
\begin{eqnarray}
\label{lagrangian3}
\mathcal{L}&\supset&-y_N^{}\bar{l}_L^{}\tilde{\phi}_L^{}R
+y_N^{}\frac{1}{M}M_N^{D}\bar{l}_L^{}\tilde{\phi}_L^{}L\nonumber\\
&&+\frac{\lambda_3^{}}{4\lambda_2^{}v_R^{}}y_N^{}\phi_L^\dagger
\phi_L^{}\bar{\nu}_R^{}L+\textrm{H.c.}\nonumber\\
&&-\frac{1}{2}M\left(\bar{L}L+\bar{R}R\right)\,.
\end{eqnarray}
Here the dimension-5 operator
$\phi^\dagger_{L}\phi^{}_L\bar{\nu}_R^{}R$ and its CP conjugate have
been omitted since they are sufficiently suppressed in comparison
with the couplings in (\ref{lagrangian3}).

Clearly, Eq. (\ref{lagrangian3}) can accommodate the standard
leptogeneisis scenario, where the 2-body decays of the Majorana
fermions into the SM lepton and Higgs doublets can generate a lepton
asymmetry if CP is not conserved. This lepton asymmetry can be
partially converted to a baryon asymmetry through sphaleron
\cite{krs1985}. So, the absence of baryonic antimatter is well
explained. The final baryon asymmetry can be described by
\begin{eqnarray}
\label{ba} \frac{n_b^{}}{s} &=&c
\Sigma_j^{}\frac{\kappa_{R_j^{}}^{}\varepsilon_{R_j^{}}^{}+\kappa_{L_j^{}}^{}\varepsilon_{L_j^{}}^{}}{g_\ast^{}}\,.
\end{eqnarray}
Here $c=-\frac{28}{79}$ \cite{krs1987} is the sphaleron induced
lepton-to-baryon conversion coefficient, $g_{\ast}^{}\simeq 112$ is
the relativistic degrees of freedom (the SM fields plus three
right-handed neutrinos), $\kappa_{R_j^{}}^{}(y_N^{},M)$ and
$\kappa_{L_j^{}}^{}(y_N^{},M)$ are wash out factors,
$\varepsilon_{R_j^{}}^{}(\varepsilon_{L_j^{}}^{})$ is the lepton
asymmetry induced by the decays of a $R_j^{}(L_j^{})$ and is
calculated at one-loop order,
\begin{subequations}
\begin{eqnarray}
\label{cpasymmetry1} \varepsilon_{R_j^{}}^{}(y_N^{},M)
&=&\frac{\Gamma_{R_j^{} \rightarrow l_L^{} +
\phi_{L}^{}}^{}-\Gamma_{R_j^{} \rightarrow l_{L}^{c} +
\phi_L^{\ast}}^{}}{\Gamma_{R_j^{}}^{}}\,,\\
\label{cpasymmetry2} \varepsilon_{L_j^{}}^{}(y_N^{},M)
&=&\frac{\Gamma_{L_j^{} \rightarrow l_L^{} +
\phi_{L}^{}}^{}-\Gamma_{L_j^{} \rightarrow l_{L}^{c} +
\phi_L^{\ast}}^{}}{\Gamma_{L_j^{}}^{}}\,,
\end{eqnarray}
\end{subequations}
where $\Gamma_{R_j^{}}^{}$ and $\Gamma_{L_j^{}}^{}$ are the decay
width at tree level,
\begin{subequations}
\begin{eqnarray}
\Gamma_{R_j^{}}^{}&=&\Gamma_{R_j^{} \rightarrow l_L^{}
\phi_{L}^{}}^{}+\Gamma_{R_j^{} \rightarrow l_{L}^{c}
\phi_L^{\ast}}^{}\nonumber\\
&=&\frac{1}{8\pi}\left(y_N^\dagger y_N^{}\right)_{jj}^{}M_j^{}\,,\\
\Gamma_{L_j^{}}^{}&=&\Gamma_{L_j^{} \rightarrow l_L^{}
\phi_{L}^{}}^{}+\Gamma_{L_j^{} \rightarrow l_{L}^{c}
\phi_L^{\ast}}^{}+\Gamma_{L_j^{}
\rightarrow\chi\phi_L^{\ast}\phi_L^{}
}^{}\nonumber\\
&=&\frac{1}{8\pi}\left(M_N^{D\dagger}\frac{1}{M}y_N^\dagger
y_N^{}\frac{1}{M}M_N^D\right)_{jj}^{}M_j^{}\nonumber\\
&&+\frac{\lambda_3^2}{2^{12}_{}\pi^3_{}\lambda_2^2}\left(y_N^\dagger
y_N^{}\right)_{jj}^{}\frac{M_j^{3}}{v_R^2}\nonumber\\
&\simeq&\mathcal{O}\left[\left(\frac{M_N^D}{M_N^M}\right)^2_{}\right]\frac{1}{8\pi}\left(y_N^\dagger
y_N^{}\right)_{jj}^{}M_j^{}\,.
\end{eqnarray}
\end{subequations}
Although the Yukawa couplings of $L$ are much smaller than those of
$R$ [cf. (\ref{lagrangian3})], the lepton asymmetries
(\ref{cpasymmetry1}) and (\ref{cpasymmetry2}) can arrive at a same
magnitude,
\begin{eqnarray}
\varepsilon_{R_j^{}}^{}\simeq \varepsilon_{L_j^{}}^{}\simeq
\varepsilon_{j}^{}\,,
\end{eqnarray}
because the Yukawa couplings of $R$ dominate the loop corrections in
the decays of both $R$ and $L$. For generating the observed baryon
asymmetry \cite{dunkley2008}
\begin{eqnarray}
\eta=\frac{n_b^{}}{n_\gamma^{}}=7.04\times \frac{n_b^{}}{s}
=(6.225\pm0.170)\times 10^{-10}_{}\,,
\end{eqnarray}
the Yukawa couplings $y_N^{}$ and the masses $M$ should fulfill
\begin{eqnarray}
\label{input1}
c\Sigma_j^{}\left(\kappa_{R_j^{}}^{}+\kappa_{L_j^{}}^{}\right)\varepsilon_j^{}\simeq
0.99\times 10^{-8}\,.
\end{eqnarray}

\section{Dark matter}

The 3-body decays of $L$ don't have significant contributions to the
total decay width and the lepton asymmetry, but it is sufficient to
produce abundant right-handed neutrinos for the dark matter relic
density. We now clarify this interesting scenario in details. It is
easy to read the relic density,
\begin{eqnarray}
\label{dmr} \frac{n_{\chi_i^{}}^{}}{s} &=&\Sigma_j^{}
\frac{\kappa'^{}_{L_{j}}Br_{ji}^{}}{g_\ast^{}}\,,
\end{eqnarray}
where $\kappa'^{}_{L_{j}^{}}(y_N^{},M)$ is a wash out factor and
\begin{eqnarray}
\label{branch}
Br_{ji}^{}&=&\frac{\Gamma_{L_j^{}\rightarrow\chi_i^{}\phi_L^\ast\phi_L^{}}^{}}
{\Gamma_{L_j^{}}^{}}\nonumber\\
&=&\mathcal{O}\left[\left(\frac{M_N^M}{M_N^D}\right)^2_{}\right]\frac{M_j^2}{v_R^2}
\frac{\lambda_3^2}{2^9_{}\pi^2_{}\lambda_2^2}\frac{\left|\left(U^\dagger_{}y_N^{}\right)_{ij}^{}\right|^2_{}}{\left(y^\dagger_{N}y_N^{}\right)_{jj}^{}}
\end{eqnarray}
is the branching ratio. From the baryon asymmetry (\ref{ba}) and the
relic density (\ref{dmr}), we can perform the following ratio,
\begin{eqnarray}
\Sigma_j^{}
\kappa_{L_j^{}}Br_{ji}^{}m_{\chi_i^{}}^{}:c\Sigma_j^{}\left(\kappa_{R_j^{}}^{}+\kappa_{L_j^{}}^{}\right)
\varepsilon_j^{}m_N^{} = \Omega_{\chi_i^{}}^{}:\Omega_b^{}\,.
\end{eqnarray}
For a given relic density, we can determine the masses of the
right-handed neutrinos by the observed baryon asymmetry,
\begin{eqnarray}
m_{\chi_i^{}}^{}=m_N^{}\frac{c\Sigma_j^{}\left(\kappa_{R_j^{}}^{}+\kappa_{L_j^{}}^{}\right)
\varepsilon_j^{}}{\Sigma_{j}^{}\kappa'^{}_{L_j^{}}Br_{ji}^{}} \frac{
\Omega_{\chi_i^{}}^{}}{\Omega_b^{}}
\end{eqnarray}
Here $m_N^{}\sim 940\,\textrm{MeV}$ is the nucleon mass. On the
other hand, as shown in Eq. (\ref{masslr}), the mass matrix of the
right-handed neutrinos has a same texture with that of the
left-handed neutrinos. For example, let's consider the case with the
degenerate neutrino masses,
\begin{eqnarray}
\label{input2} m_{1}^{}\simeq m_{2}^{}\simeq
m_{3}^{}=0.05\,\textrm{eV}\,,
\end{eqnarray}
which immediately yields
\begin{eqnarray}
\label{input3} m_{\chi_{1}^{}}^{}\simeq m_{\chi_{2}^{}}^{}\simeq
m_{\chi_{3}^{}}^{}=0.05\,\textrm{eV}\frac{v_L^2}{v_R^2}\,.
\end{eqnarray}
With further assumptions,
\begin{eqnarray}
\kappa^{}_{L_j^{}}\simeq\kappa^{}_{R_j^{}}\simeq
\kappa'^{}_{L_j^{}}=0.1\,,
\end{eqnarray}
which is true in the weak washout region, and
\begin{eqnarray}
\label{input4} \frac{\lambda_3^{}}{\lambda_2^{}}=3.3\times
10^{-4}_{},~\frac{M_N^M}{M_N^D}=1.9\times 10^7_{}\,,~
\frac{M_j^{}}{v_R^{}}=2.1\times 10^{-4}_{}\,,\nonumber
\end{eqnarray}
\vspace{-5mm}
\begin{eqnarray}
\end{eqnarray}
we can obtain the dark matter masses
\begin{eqnarray}
\label{input5} m_{\chi_{1,2,3}^{}}^{}\simeq 1.3\,\textrm{MeV}\,.
\end{eqnarray}
Accordingly, the left-right symmetry breaking scale can be given by
\begin{eqnarray}
\label{input6}
v_R^{}=v_L^{}\sqrt{\frac{m_{\chi_i^{}}^{}}{m_i^{}}}\simeq 5.1\times
10^3_{}\,v_L^{}\,.
\end{eqnarray}
The above breaking scale points to the necessity of the resonant
\cite{fps1995,pilaftsis1997} leptogenesis because of the parameter
choice (\ref{input4}), i.e. $M_j^{}\simeq 186\,\textrm{GeV}$. It
also indicates that the gauge interactions of the right-handed
neutrinos decouple at a high temperature
$\sim\left(\frac{v_R^{}}{v_L^{}}\right)^{\frac{4}{3}}_{}
\mathcal{O}(\textrm{MeV})=\mathcal{O}(100\,\textrm{GeV})$ so that
the non-thermally \cite{lhzb2000} produced right-handed neutrinos
can successfully explain the dark matter relic density.

We should keep in mind that the right-handed neutrinos mix with
their left-handed partners due to the Dirac mass term in Eq.
(\ref{lagrangian2}) so that they will decay at tree level and loop
orders. In absence of the mixing between $W_L^{\pm}$ and $W_R^{\pm}$
(cf. (\ref{wmass})), the decay width should be \cite{pw1982,bhl2009}
\begin{subequations}
\label{dmd}
\begin{eqnarray}
\label{dmd1} \Gamma_{\chi_i^{}\rightarrow  \nu\nu\nu}^{}&=&\frac{
G_F^2}{384\pi^3}\sin^2_{}\left(2\theta_i^{}\right)m_{\chi_i^{}}^5\,,\\
\label{dmd2} \Gamma_{\chi_i^{}\rightarrow  \nu
e^+_{}e^-_{}}^{}&=&\frac{5
G_F^2}{3072\pi^3}\sin^2_{}\left(2\theta_i^{}\right)m_{\chi_i^{}}^5\,,\\
\label{dmd3} \Gamma_{\chi_i^{}\rightarrow
\nu\gamma}^{}&=&\frac{9\alpha
G_F^2}{1024\pi^4}\sin^2_{}\left(2\theta_i^{}\right)m_{\chi_i^{}}^5\,.
\end{eqnarray}
\end{subequations}
Here $\alpha$ and $G_F^{}$, respectively, are the fine-structure
constant and the Fermi constant, $\theta_i^2$ is the mixing angle
defined by
\begin{eqnarray}
\theta_i^2=\frac{\left(m_D^\dagger
m_D^{}\right)_{ii}^{}}{m_{\chi_{i}^{}}^2}=
\mathcal{O}\left[\left(\frac{M_N^D}{M_N^M}\right)^2_{}\right]\frac{v_L^2}{v_R^2}\,.
\end{eqnarray}
With the previous parameter choice, we can determine the mixing
angle to be $\theta_i^2\simeq 10^{-22}_{}$. Therefore the decay into
the electron-positron pairs (\ref{dmd2}) can \cite{pp2004} provide a
natural explanation for the flux of $511\,\textrm{keV}$ photons from
the galactic bulge observed by INTEGRAL \cite{jean2003} satellite,
\begin{eqnarray}
\frac{\Phi_{511\gamma}^{}}{\Phi_{\textrm{exp}}^{}}\simeq\Sigma_i^{}
\frac{\theta_i^{2}}{
10^{-22}_{}}\left(\frac{m_{\chi_i^{}}}{1.3\,\textrm{MeV}}\right)^4_{}\frac{\Omega_{\chi_i^{}}^{}}{\Omega_{\chi}^{}}\,.
\end{eqnarray}
It is easy to check that our scenario is consistent with other
astrophysical and cosmological constraints \cite{brs2009}.
Alternatively, we may explain the observed cosmic positron/electron
excess
\cite{chang2008,torii2008,adriani2008,aharonian2008,abdo2009}, which
is probably from continuum distribution of pulsars
\cite{profumo2008,bgkms2009}, by fine tuning the parameters.

\section{Conclusion}

In summary we have shown the dark matter can be well determined by
the neutrino masses and the baryon asymmetry in the left-right
symmetric model with doublet and singlet fields. In this model, the
SM fermions obtain masses by integrating out charged singlet
fermions. In the neutrino sector, the right(left)-handed neutral
fermions, associated with the left(right)-handed Higgs doublet, can
generate the left(right)-handed neutrino masses through the seesaw.
The mass matrices of the left- and right-handed neutrinos have a
same structure as a result of the left-right symmetry. The neutral
singlets are also responsible for the baryon asymmetry and the dark
matter. Specifically their 2-body decays can produce a desired
lepton asymmetry in the left-handed leptons and then the observed
baryon asymmetry can be realized by the sphaleron induced
lepton-to-baryon conversion. At the same time, the right-handed
neutrinos can serve as the dark matter as they have a right relic
density from the 3-body decays of the neutral singlets. The decays
of these non-thermally produced right-handed neutrinos can easily
induce the observed fluxes of 511 keV photons from the Galactic
bulge. The attractive feature of our scenario is that the left-right
symmetry can connect the properties of the dark matter to the
neutrino masses and the baryon asymmetry.

\vspace{2mm}

\textbf{Acknowledgement}:  I thank Manfred Lindner for hospitality
at Max-Plank-Institut f\"{u}r Kernphysik. This work is supported by
the Alexander von Humboldt Foundation.


\begin{thebibliography}{99}


\bibitem{dunkley2008}
J. Dunkley {\it et al.}, [WMAP Collaboration], Astrophys. J. Suppl.
\textbf{180}, 306 (2009).


\bibitem{kuzmin1997}
V.A. Kuzmin, arXiv: hep-ph/9701269; R. Kitano and I. Low, Phys. Rev.
D \textbf{71}, 023510 (2005); N. Cosme, L. Lopez Honorez, and M.H.G.
Tytgat, Phys. Rev. D \textbf{72}, 043505 (2005); D.E. Kaplan, M.A.
Luty, and K.M. Zurek, Phys. Rev. D \textbf{79}, 115016 (2009); P.H.
Gu, U. Sarkar, and X. Zhang, Phys. Rev. D \textbf{80}, 076003
(2009); I.M. Shoemaker and A. Kusenko, Phys. Rev. D \textbf{80},
075021 (2009); P.H. Gu and U. Sarkar, arXiv:0909.5463 [hep-ph]; H.
An, S.L. Chen, R.N. Mohapatra, and Y. Zhang, arXiv:0911.4463
[hep-ph].





\bibitem{stv2008}
T. Schwetz, M.A. T\'{o}rtola, J.W.F. Valle, New J. Phys.
\textbf{10}, 113011 (2008).


\bibitem{minkowski1977}
P. Minkowski, Phys. Lett. B \textbf{67}, 421 (1977).

\bibitem{yanagida1979}
T. Yanagida, in {\it Proc. of the Workshop on Unified Theory and the
Baryon Number of the Universe}, ed. O. Sawada and A. Sugamoto (KEK,
Tsukuba, 1979), p. 95.

\bibitem{grs1979}
M. Gell-Mann, P. Ramond, and R. Slansky, in {\it Supergravity}, ed.
F. van Nieuwenhuizen and D. Freedman (North Holland, Amsterdam,
1979), p. 315.

\bibitem{glashow1980}
S.L. Glashow, in {\it Quarks and Leptons}, ed. M. L\'{e}vy {\it et
al.} (Plenum, New York, 1980), p. 707.

\bibitem{ms1980}
R.N. Mohapatra and G. Senjanovi\'{c}, Phys. Rev. Lett. \textbf{44},
912 (1980).


\bibitem{mw1980}
M. Magg and C. Wetterich, Phys. Lett. B \textbf{94}, 61 (1980); J.
Schechter and J.W.F. Valle, Phys. Rev. D \textbf{22}, 2227 (1980);
T.P. Cheng and L.F. Li, Phys. Rev. D \textbf{22}, 2860 (1980); G.
Lazarides, Q. Shafi, and C. Wetterich, Nucl. Phys. B \textbf{181},
287 (1981); R.N. Mohapatra and G. Senjanovi\'{c}, Phys. Rev. D
\textbf{23}, 165 (1981).




\bibitem{fy1986}
M. Fukugita and T. Yanagida, Phys. Lett. B \textbf{174}, 45 (1986).


\bibitem{lpy1986}
P. Langacker, R.D. Peccei, and T. Yanagida, Mod. Phys. Lett. A
\textbf{1}, 541 (1986); M.A. Luty, Phys. Rev. D \textbf{45}, 455
(1992); R.N. Mohapatra and X. Zhang, Phys. Rev. D \textbf{46}, 5331
(1992).



\bibitem{fps1995}
M. Flanz, E.A. Paschos, and U. Sarkar, Phys. Lett. B \textbf{345},
248 (1995); M. Flanz, E.A. Paschos, U. Sarkar, and J. Weiss, Phys.
Lett. B \textbf{389}, 693 (1996); L. Covi, E. Roulet, and F.
Vissani, Phys. Lett. B \textbf{384}, 169 (1996).





\bibitem{pilaftsis1997}
A. Pilaftsis, Phys. Rev. D \textbf{56}, 5431 (1997).

\bibitem{ms1998}
E Ma and U. Sarkar, Phys. Rev. Lett. \textbf{80}, 5716 (1998).


\bibitem{bpy2005}
W. Buchm\"{u}ller, R.D. Peccei, and T. Yanagida, Ann. Rev. Nucl.
Part. Sci. \textbf{55}, 311 (2005), and references therein.

\bibitem{ps1974}
J.C. Pati and A. Salam, Phys. Rev. D \textbf{10}, 275 (1974); R.N.
Mohapatra and J.C. Pati, Phys. Rev. D \textbf{11}, 566 (1975); R.N.
Mohapatra and J.C. Pati, Phys. Rev. D \textbf{11}, 2558 (1975); R.N.
Mohapatra and G. Senjanovi\'{c}, Phys. Rev. D \textbf{12}, 1502
(1975).


\bibitem{bmu2003}
B. Brahmachari, E Ma, and U. Sarkar, Phys. Rev. Lett. \textbf{91},
011801 (2003).


\bibitem{siringo2003}
F. Siringo, Eur. Phys. J. C \textbf{32}, 555 (2004).

\bibitem{cmp1984}
D. Chang, R.N. Mohapatra, and M.K. Parida, Phys. Rev. Lett.
\textbf{52}, 1072 (1984); D. Chang, R.N. Mohapatra, and M.K. Parida,
Phys. Rev. D \textbf{30}, 1052 (1984).



\bibitem{mns1962}
Z. Maki, M. Nakagawa, and S. Sakata, Prog. Theor. Phys. \textbf{28},
870 (1962).


\bibitem{krs1985}
V.A. Kuzmin, V.A. Rubakov, and M.E. Shaposhnikov, Phys. Lett. B
\textbf{155}, 36 (1985).


\bibitem{krs1987}
V.A. Kuzmin, V.A. Rubakov, and M.E. Shaposhnikov, Phys. Lett. B
\textbf{191}, 171 (1987); M. Shaposhnikov, Nucl. Phys. B
\textbf{308}, 885 (1988).




\bibitem{lhzb2000}
W.B. Lin, D.H. Huang, X. Zhang, and R.H. Brandenberger, Phys. Rev.
Lett. \textbf{86}, 954 (2001).

\bibitem{pw1982}
P.B. Pal and L. Wolfenstein, Phys. Rev. D \textbf{25}, 766 (1982);
V. Barger, R.J.N. Phillips, and S. Sarkar, Phys. Lett. B
\textbf{352}, 365 (1995).


\bibitem{bhl2009}
F. Bezrukov, H. Hettmansperger, and M. Lindner, arXiv:0912.4415
[hep-ph].





\bibitem{pp2004}
C. Picciotto and M. Pospelov, Phys. Lett. B \textbf{605}, 15 (2005);
S. Khalil and O. Seto, JCAP \textbf{0810}, 024 (2008).



\bibitem{jean2003}
P. Jean {\it et al.}, Astron. Astrophys. \textbf{407}, L55 (2003).



\bibitem{brs2009}
A. Boyarsky, O. Ruchayskiy, and M. Shaposhnikov, Ann. Rev. Nucl.
Part. Sci. \textbf{59}, 191 (2009), and references therein.






\bibitem{chang2008}
J. Chang, {\it et al.}, [ATIC Collaboration], Nature \textbf{456},
362 (2008).


\bibitem{torii2008}
S. Torii, {\it et al.}, [PPB-BETS Collaboration], arXiv: 0809.0760
[astro-ph].


\bibitem{adriani2008}
O. Adriani, {\it et al.}, [PAMELA Collaboration], Nature
\textbf{458}, 607 (2009).


\bibitem{aharonian2008}
F.\ Aharonian {\it et al.}, [H.E.S.S.\ Collaboration], Astron.
Astrophys. \textbf{508:}, 561 (2009).


\bibitem{abdo2009}
A.A. Abdo, {\it et al.}, [Fermi/LAT Collaboration], Phys. Rev. Lett.
\textbf{102}, 181101 (2009).

\bibitem{profumo2008}
S. Profumo, arXiv:0812.4457 [astro-ph].


\bibitem{bgkms2009}
V. Barger, Y. Gao, W.Y. Keung, D. Marfatia, and G. Shaughnessy,
Phys. Lett. B \textbf{678}, 283 (2009).


\end{thebibliography}
\end{document}